%% file: paper_SJ-CP-MW.tex
\documentclass[12pt]{article}
\usepackage{ametsoc}
\usepackage{subfigure}

\usepackage{color}

\newcommand{\be}{\begin{equation}}
\newcommand{\ee}{\end{equation}}
\newcommand{\beq}{\begin{eqnarray*}}
\newcommand{\eeq}{\end{eqnarray*}}

\newcommand{\myabstract}{
Extreme event attribution characterizes how anthropogenic climate change may have influenced the probability and magnitude of selected individual extreme weather and climate events. Attribution statements often involve quantification of the fraction of attributable risk (FAR) or the risk ratio (RR) and associated confidence intervals. Many such analyses use climate model output to characterize extreme event behavior with and without anthropogenic influence. However, such climate models may have biases in their representation of extreme events. To account for discrepancies in the probabilities of extreme events between observational datasets and model datasets, we demonstrate an appropriate rescaling of the model output based on the quantiles of the datasets to estimate an adjusted risk ratio. Our methodology accounts for various components of uncertainty in estimation of the risk ratio. In particular, we present an approach to construct a one-sided confidence interval on the lower bound of the risk ratio when the estimated risk ratio is infinity. 
We demonstrate the methodology using the summer 2011 central US heatwave and output from the Community Earth System Model. In this example, we find that the lower bound of the risk ratio is relatively insensitive to the magnitude and probability of the actual event.
}
\begin{document}
%
%
\title{\textbf{\large{Quantile-based Bias Correction and Uncertainty Quantification of Extreme Event Attribution Statements}}}
%
%
\author{Soyoung Jeon\\Earth Sciences Division, Lawrence Berkeley National Laboratory\\Berkeley, California, USA\\soyoungj82@gmail.com\\ \\
Christopher J. Paciorek\\Department of Statistics, University of California\\Berkeley, California, USA\\paciorek@stat.berkeley.edu\\ \\
Michael F. Wehner\\Computational Research Division, Lawrence Berkeley National Laboratory\\Berkeley, California, USA\\mfwehner@lbl.gov}

%
\ifthenelse{\boolean{dc}}
{
\twocolumn[
\begin{@twocolumnfalse}
\amstitle

\begin{center}
\begin{minipage}{13.0cm}
\begin{abstract}
	\myabstract
	\newline
	\begin{center}
		\rule{38mm}{0.2mm}
	\end{center}
\end{abstract}
\end{minipage}
\end{center}
\end{@twocolumnfalse}
]
}
{
\amstitle
\begin{abstract}
\myabstract
\end{abstract}
\newpage
}

%

\section{Introduction}
\input{introduction} \label{s:intro}

\section{Case Study: Summer 2011 Central USA Heatwave} \label{s:data}
\input{data}

\section{Methodology}
\input{method}


\section{Results} \label{s:results}
\input{result}

\section{Conclusion}
\input{conclusion}

\input{figures_and_tables}

%

\begin{acknowledgment} 
This research was supported by the Director, Office of Science, Office of Biological and Environmental Research of the U.S. Department of Energy under Contract No. DE-AC02-05CH11231 as part of their Regional and Global Climate Modeling Program and used resources of the National Energy Research Scientific Computing Center (NERSC), also supported by the Office of Science of the U.S. Department of Energy, under Contract No. DE-AC02-05CH11231. This document was prepared as an account of work sponsored by the United States Government. While this document is believed to contain correct information, neither the United States Government nor any agency thereof, nor the Regents of the University of California, nor any of their employees, makes any warranty, express or implied, or assumes any legal responsibility for the accuracy, completeness, or usefulness of any information, apparatus, product, or process disclosed, or represents that its use would not infringe privately owned rights. Reference herein to any specific commercial product, process, or service by its trade name, trademark, manufacturer, or otherwise, does not necessarily constitute or imply its endorsement, recommendation, or favoring by the United States Government or any agency thereof, or the Regents of the University of California. The views and opinions of authors expressed herein do not necessarily state or reflect those of the United States Government or any agency thereof or the Regents of the University of California.
\end{acknowledgment}

\ifthenelse{\boolean{dc}}
{}
{\clearpage}
\begin{appendix}
\section*{\begin{center}Background for Modeling of Extreme Values \end{center}}
\input{appendix}

%
%
%
%
%
%
\end{appendix}

\ifthenelse{\boolean{dc}}
{}
{\clearpage}
\bibliographystyle{ametsoc}
\bibliography{references0}

\end{document}

%% file: introduction.tex

The summer of 2011 was extremely hot in Texas and Oklahoma, producing a record of 30.26$^{\circ}$C for the average June-July-August (JJA) temperature (3.24$^{\circ}$C above the 1961-1990 mean) as measured in the CRU observational dataset (CRU TS 3.21, \cite{Harris_et_al:2014}). In a previous study of the 2011 Texas heat wave by \cite{Hoerling_et_al:2008}, a major factor contributing to the magnitude of 2011 heat wave was the severe drought over Texas resulting from the La Ni\~{n}a phase of the ocean state. However, the analysis found a substantial anthropogenic increase in the chance of an event of this magnitude. As in most mid-latitude land regions, the probability of extreme summer heat in this region has increased due to human-induced climate change \citep{Min_et_al:2013}. However, as \cite{Stone_et_al:2013} note, depending on spatial extent of the region analyzed, observed summer warming is low in Texas in 2011 and traceable to the so-called ``warming hole" \citep{Meehl_et_al:2012}.


 
Extreme event attribution analyses attempt to characterize whether and how the probability of an extreme event has changed because of external forcing, usually anthropogenic, of the climate system. As with traditional detection and attribution of trends in climate variables \citep{Bindoff:2013}, climate models must play an important role in the methodology due to the absence of extremely long observational records. The fraction of attributable risk ($FAR$) or the risk ratio ($RR$) are  commonly-used measures that quantify this potential human influence \citep{Palmer:1999, Allen:2003, Stott_et_al:2004, Jaeger_et_al:2008, Pall_et_al:2011, Wolski_et_al:2013}. Following the notation used in \cite{Stott_et_al:2004}, let $p_{A}$ be the probability in a simulation using all external (anthropogenic plus natural) forcings of an event of similar magnitude, location and season to the actual event and $p_{C}$ be the probability of such an event under natural forcings. The $FAR$ is defined as $FAR=1-p_C/p_A$ while the $RR$ is defined as $RR=p_A/p_C$, with each quantity a simple mathematical transformation of the other. We note that the commonly used term ``risk ratio" is more precisely a ``probability ratio" \citep{Fischer_Knutti:2015} but we will stick to the $RR$ nomenclature in this study---in part because $RR$ is well-established terminology.

In the seminal study of the 2003 European heat wave by \cite{Stott_et_al:2004}, their climate model did remarkably well in simulating both European mean summer temperature and its interannual standard deviation. However, this is not generally the case for the entirety of available climate model outputs nor for the wide range of extreme events of current interest \citep{BAMS:2011, BAMS:2012, BAMS:2013}. Hence there is a need to correct model output, particularly in the tail of its distribution, to more realistically estimate both $p_{A}$ and $p_{C}$. Quantile-based mapping is often used to reduce such climate model biases in statistical downscaling studies of future climate change projections. Such methods match quantiles of climate model outputs to observed data for monthly GCM temperature and precipitation \citep{Wood_et_al:2004}. For instance, quantile-based corrections to the transfer function between the coarse mesh of the global models and the finer downscaled mesh have been obtained by using cumulative distribution functions (CDFs) to match percentiles between the model outputs and observations over a specified base period \citep{Maurer_et_al:2008}. \cite{Li_et_al:2010} proposed an adjustment of the traditional quantile matching method \citep{Panofsky_Brier:1968} to account for time-dependent changes in the distribution of the future climate and suggested that the quantile-matching method is a simple and straightforward method for reducing the scale differences between simulations and observations, for the tails of the distribution as well. The quantile mapping approach 
of \cite{Li_et_al:2010} has been previously used to empirically estimate annual and decadal maximum daily precipitation in an attribution study of an early season blizzard in western South Dakota \citep{Edwards_et_al:2014}.

This paper is concerned with developing a formal statistical methodology using extreme value analysis combined with quantile mapping to adjust for model  biases in event attribution analyses. We apply the methodology to the 2011 central US heatwave as a case study, using an ensemble of climate model simulations. In Section 2, we describe the observed and simulated data for the central US heatwave analysis. Section 3 presents our statistical methodology, describing the use of extreme value methods combined with the quantile bias correction to estimate the risk ratio. We describe several approaches for estimating uncertainty in the risk ratio, focusing on the use of a likelihood ratio-based confidence interval that provides a one-sided interval even when the estimated risk ratio is infinity. 
In Section 4 we present results from using the methodology for event attribution for the central US heatwave, showing strong evidence of anthropogenic influence.

%% file: data.tex


For a representative case study of extreme temperature attribution, we define a  central United States region bordered by 90$^{\circ}$W to 105$^{\circ}$W in longitude and 25$^{\circ}$N to 45$^{\circ}$N in latitude, chosen to encompass the Texas and Oklahoma heatwave that occurred in summer 2011 (see Figure \ref{f:map}). For this region, we calculated summer (June, July, August [JJA]) average temperature anomalies for the time period 1901-2012 by averaging daily maximum temperatures for grid cells falling within the study region. Anomalies are computed using 1961-1990 as the reference period. 

The observational data in this study are obtained from the gridded data product (CRU TS 3.21, Climatic Research Unit Time Series) available on a 0.5$^{\circ} \times$0.5$^{\circ}$ grid provided by the Climatic Research Unit \citep{Harris_et_al:2014}. This dataset provides monthly average daily maximum surface air temperature anomalies. 
Similarly, monthly averaged daily maximum surface air temperatures were obtained from the CMIP5 database through the Earth System Grid Federation (ESGF) archive. For both the observations and model output, spatial averages over the cells covering the land surface of the region were calculated, resulting in simple 1-dimensional time series. In this study, we use a single climate model, the fourth version of the Community Climate System Model (CCSM4) with a resolution of 1.25$^{\circ}\times$0.94$^{\circ}$ grid. To more fully explore the structural uncertainty in event attribution statements, additional models would need to be included in the analysis. While that topic is outside the scope of this paper, our methodology is also relevant for analyses that use multiple models that will each have their own biases.

The CCSM4 ensemble consists of multiple  simulations, each initialized from different times of a control run; we treat the ensemble members as independent realizations of the model's possible climate state. For the actual scenario with all forcings included, we use an ensemble of five members, constructed by concatenating the period 1901-2005 from the CMIP5 ``historical" forcings experiment and the period 2006-2012 from the matching RCP8.5 emissions scenario experiment. As a representation of a world without human interference on the climate system, we construct a counterfactual scenario by producing an ensemble of 12 100-year segments drawn from the preindustrial control run. In this scenario, greenhouse gases, aerosols and stratospheric ozone concentrations are set at pre-industrial levels, but other external natural forcings such as solar variability and volcanoes are not included. We use this counterfactual scenario as a proxy for the natural climate system without any external forcing factors.

An important consideration in event attribution analyses is whether the climate model(s) reasonably represent the magnitudes and frequencies of the event of interest \citep{Chrisidis_et_al:2013}. 
Figure \ref{f:ts_all} shows that summer temperatures vary more in the CCSM4 output than in the observations. The record observed extreme value in our central US region in 2011 was 2.467$^{\circ}$C above the 1961-1990 average (represented by the large black dot); even this extreme is somewhat lower than the observed values over just the states of Texas and Oklahoma. However, this value is not particularly rare in either model scenario dataset. Due to this scale mismatch in temperature variability, the climate model incorrectly estimates the probabilities of extreme events of this magnitude in both scenarios. In light of this model bias, a quantile mapping procedure to scale the extreme values of either the model or the observations to the other is warranted to more consistently relate the model's risk ratio to the real world. More precisely, we define the event according to observations, even in the presence of observational error, and calibrate the model to the observations with the quantile-based method described in this paper. 
The methodology presented in Section 3 implements such a scaling by first estimating the probability, $\hat p_O$, of reaching or exceeding the actual event magnitude from the observations. Then, the magnitude, $\hat z_A$, of an event in that time with the same  probability, $\hat p_O(=\hat p_A)$, is estimated from the actual scenario of the model. The risk ratio can then be estimated from the probability, $\hat p_C$, of an event of magnitude $\hat z_A$ from the counterfactual scenario of the model as  $\widehat{RR}=\hat p_A/\hat p_C$. 
   
Implicit in this estimation of $RR$ is an assumption that the asymptotic behavior of the all forcings model ensemble is similar to the observations. Indeed, it is not clear how to validate that assumption given the limited observational data availability and the rarity of the events of interest in attribution studies. However, it is clear that errors from estimating $RR$ directly from the model without a quantile mapping correction would be larger, because probability estimates would be drawn from a different part of the distribution. In this case study, such probabilities would not be representative of the tail of the distribution. Furthermore, in other cases, the model may underestimate variability, and the probability in the model of an event of the actual magnitude may be zero due to the boundedness of the distribution function. We return to the implications of bounded distributions for uncertainty estimates in Section 3. There is a risk that bias correction could mask serious model errors in simulating the processes responsible for the extreme event in question. This risk is also present in more commonly-used bias correction techniques such as the use of anomalies based on subtracting off or dividing by a reference value. In the present example, a complete assessment of the robustness of the results would also include analysis of CCSM4's ability to reproduce the type of large-scale meteorological patterns leading to central US heatwaves as well as its simulation of ENSO.

%% file: method.tex

\subsection{Quantile Bias Correction}


Here we describe a quantile mapping methodology to adjust for the difference in scales between observations and model outputs; we call this methodology \textit{quantile bias correction}. The methodology seeks to estimate adjusted probabilities $p_A$ and $p_C$ and the corresponding $RR$. From this point forward, since we will work exclusively with the adjusted probabilities, we will simply use $p_A$ and $p_C$ to refer to the adjusted probabilities rather than introduce additional notation to distinguish adjusted and unadjusted probabilities.  The steps of the method are as follows:
\begin{itemize}
\item[(1)] observe some extreme event, e.g., the extreme value of 2.467$^{\circ}$C for the 2011 central US heatwave, and  estimate the probability, $\hat{p}_O$, of the observed event using appropriate extreme value statistical methods, 
\item[(2)] use extreme value methods applied to the model output under the actual scenario to estimate the magnitude, $\hat{z}_A$, associated with the probability $\hat{p}_O$, thereby defining $p_A=p_O$,   
\item[(3)] use extreme value methods applied to the model output under the counterfactual scenario to estimate the probability $\hat{p}_{C}$ of exceeding the value $\hat{z}_A$, and 
\item[(4)] calculate the estimated risk ratio $\widehat{RR}={\hat {p}_A / {\hat {p}_C}}$.    
\end{itemize}
Step 2 is the critical bias adjustment, where the method adjusts the magnitude of the extreme event considered in the model output to be of the same rarity in the model under the all forcings scenario as the actual extreme event is in the observations. This correction in the tail of the distribution is likely to be very different than a simple adjustment of the model mean and/or variance and more appropriate to event attribution studies. Figure \ref{f:CDFs} illustrates the quantile bias correction method and demonstrates the steps with cumulative distribution functions for the 2011 central US heatwave analysis.

\subsection{Using Extreme Value Statistics to Estimate Event Probabilities}

The probabilities, $p_O$ and $p_C$, can be estimated using a variety of techniques. For instance, in studies using ensembles with tens of thousands of model realizations \citep{Pall_et_al:2011}, probabilities of very rare events can often be estimated simply using the proportion of realizations in which the event was observed. However, in our case study, as will be the case in many other analyses, there are only a few simulations and the tail of the distribution is not well sampled.  Extreme value statistical methods involve fitting a three parameter extreme value distribution function to a subset of the available sample and are well suited to estimating such probabilities. After estimating the distribution's  parameters, step 2 can be accomplished by inverting the distribution to estimate the magnitude of $\hat{z}_A$ in the form of a return value for the period $1/{\hat{p}_O}$. 

In the current study, we use a point process (PP) approach to extreme value analysis \citep{Smith:1989, Coles:2001, Katz:et:al2002, Furrer_et_al:2010}. This approach involves modeling exceedances over a high threshold and is described in detail in the Appendix. The simplest formulations of extreme value models assume that the distribution of the extremes does not change over time, an assumption of stationarity. The PP approach can be extended to non-stationary cases in which the parameters of the model, $\mu$, $\sigma$, and $\xi$, are allowed to be (arbitrary but often linear) functions of covariates. Covariates are chosen to incorporate additional physical insight into the statistical model. A common practice is to represent nonstationarity through only the location parameter, $\mu$, and take $\sigma$ and $\xi$ to be constant \citep{Coles:2001, Kharin_Zwiers:2005}. For example, one could represent the location of the extreme value distribution $\mu_t$ to depend on time $t$ as a function of time-varying covariates $x_{kt}$:

	\be
	\mu_t=\beta_0 + \sum_{k=1}^K \beta_k x_{kt}.
	\ee

The model under the actual scenario, as seen in Figure \ref{f:ts_all}, is non-stationary due to the effects of anthropogenic climate change. Rather than try to directly develop a covariate as an explicitly nonlinear function of time, it is simpler to use a more physically-based ``covariate" as a linear source of non-stationarity. A simple choice is a temporally-smoothed global mean temperature anomaly ($x_t$). A 13-point filter \citep{IPCC_AR4} removes some of the natural modes of variability that may affect central US summer temperature but retains the anthropogenic warming signal. This function is then a non-linear proxy for time that we can use as a covariate in a linear representation of the location parameter, $\mu_t=\beta_0 +\beta_1 x_t$. We note that adding additional covariates to account for other known physical dependencies, such as an El Ni\~{n}o/La Ni\~{n}a index, may improve the quality of the fitted distribution but as such is outside the scope of this study. Finally, as the model under the counterfactual scenario is presumed to be stationary, we do not use a covariate in fitting that dataset. In this study, we computed the Akaike Information Criterion (AIC) to compare stationary and non-stationary models for the observations and actual scenario output, where the model with the smaller AIC value is preferred. For the actual scenario, the non-stationary model was strongly preferred based on AIC. However, we found that the AIC for the stationary model for observations (152.93) was slightly 
smaller than the AIC for the non-stationary model for the observations (154.14). This is a consequence of the very small observed warming trend in the selected region. Despite this preference for omitting the covariate, we use the non-stationary model for the observational data to be consistent with the statistical representation for the actual scenario output.  

The PP model requires the choice of an arbitrary threshold, with only data above the threshold used to fit the model, as described in the Appendix. There are few rigid guidelines for how high the threshold should be. It must be high enough to be in the `asymptotic' regime, i.e., that the assumptions of the extreme value statistical theory are satisfied, but low enough that enough points from the original sample are retained to reduce the uncertainty in estimating the parameters of the statistical model. Here we use the 80th percentile of the values in each dataset. Standard diagnostics \citep{Coles:2001, Scarrott_MacDonald:2012}, including mean residual life plots shown in Figure \ref{f:mrlp}, suggest this is a reasonable choice. 

Given the choice of a threshold and covariates, the PP-based extreme value distribution is straightforward to fit using maximum likelihood methods, providing estimates of $\mu_t$ (i.e., ${\beta}_0$ and  ${\beta}_1$), ${\sigma}$, and ${\xi}$. To fit the model, we use the \texttt{fevd} routine of the \texttt{R} package, \texttt{extRemes} \citep{extRemes:2011}. Note that for seasonal data such as for this case study, the \texttt{time.units} argument should be specified to be \texttt{"m/year"}, where \texttt{m} is the number of observations in each block of data. It is useful to treat a `block' as a year so that return levels can be considered to be the value exceeded once in $1/p$ years. When using an ensemble of model runs, we have multiple replicates for each year, so \texttt{m} is the number of ensemble members (e.g., $\texttt{m}=5$ for the all forcings ensemble). To implement steps 2 and 3 of the quantile bias correction method, we need to be able to calculate both a return level given a specified probability $\hat{z}_A(\hat{p}_O)$ and a probability given a specified return level, $\hat{p}_C(\hat{z}_A)$. Both of these values are obtained from the estimated parameter values as shown in the Appendix, equations (\ref{eqn:return_prob}) and (\ref{eqn:return_value}).

\subsection{Uncertainty Quantification of the Risk Ratio}

We have presented an approach to estimating the $RR$ using the quantile bias correction method. We turn now to accounting for the various sources of uncertainties in the estimate of $RR$ produced by this method. Here we focus on uncertainty from statistical estimation of the various probabilities; structural uncertainty that arises from using model simulations in place of the real climate system is of course important but is beyond the scope of our work. More precisely, the uncertainties in estimating the risk ratio can be separated into three sources: uncertainty in estimating $p_O$ using the observations (step 1), uncertainty in estimating ${z}_A$ using the actual scenario model output (step 2), and uncertainty in estimating $p_C$ using the counterfactual scenario output (step 3). In this section we quantify the uncertainty in the risk ratio considering the second and third sources of uncertainty. With regard to the first source, for now we consider the magnitude of the extreme event to be a given, as a precise estimate of $p_O$ will be shown to not be absolutely necessary to make a confident attribution statement. Rather, we believe the sensitivity of the estimate of $RR$ to a defensible range of $z_O$ values (and $p_O$) is critical to confident extreme event attribution.

In our uncertainty analysis below, we condense our notation of the fitted extreme value distributions to $\theta_A = (\beta_{0_A},\beta_{1_A},\sigma_A,\xi_A)$ and $\theta_N = (\mu_{C},\sigma_C,\xi_C)$, where $A$ again indicates the model under the actual scenario and $C$ the model under the counterfactual scenario. We consider several approaches to deriving a confidence interval for the $RR$. Given that the $RR$ is non-negative and its sampling distribution is likely to be skewed, we work on the base-2 logarithmic scale.

A standard approach to estimating the standard error of a non-linear functional of parameters in a statistical model is to use the delta method and then derive a confidence interval using a normal approximation (Sections 5.5.4 \& 10.4.1, \cite{Case:Berg:2002}). Another possibility is to use the bootstrap to either estimate the standard error or directly estimate a confidence interval (Section 10.1.4, \cite{Case:Berg:2002}). However, both of these methods fail when the estimated $RR$ is infinity. The bootstrap uncertainty estimate will also pose difficulties if some of the bootstrap datasets produce estimated risk ratios that are infinity. This outcome is quite likely if the extreme value distribution of the model output under the counterfactual scenario is bounded and the magnitude of $\hat{z}_A$ is close to that bound. Therefore, after a brief discussion of the delta method and the bootstrap, we develop an alternative confidence interval by inverting a likelihood ratio test (LRT) and propose this is as a general approach to estimating a lower bound of $RR$. 
 
\paragraph{Delta Method} In this uncertainty analysis, we estimate the log risk ratio and $\log RR=f(\theta)$ as a function of the parameter vector $\theta=(\theta_A, \theta_C)$. The delta method uses an analytic approximation by a first-order Taylor series expansion: $f(\hat \theta) \approx f(\theta)+\triangledown f(\theta)^T (\hat \theta-\theta)$, where $\triangledown f$ is a vector of the partial derivatives of $f$ and $\hat \theta$ is the maximum likelihood estimate of $\theta$.
Taking the variance of both sides of the Taylor approximation above, the delta method gives that 
	\be
	\widehat{\mathrm{Var}} (\log \widehat{RR})=\widehat{\mathrm{Var}} [f(\hat \theta)] \approx \triangledown f(\hat \theta)^T \mathrm{Cov}(\hat \theta) \triangledown f(\hat \theta).
	\ee
The variance-covariance matrix of $\hat \theta$, $\mathrm{Cov}(\hat \theta)$ is based on the matrix of second derivatives of the likelihood function. The standard error is $\mathrm{s.e.}(\log \widehat{RR})=\sqrt{\widehat{\mathrm{Var}}(\log \widehat{RR})}$ and the corresponding 95\% confidence interval for $\log RR$ is 
	\be
	\big( \log \widehat{RR}-1.96\mbox{ }\mathrm{s.e.}(\log \widehat{RR}), \mbox{ } \log\widehat{RR}+1.96\mbox{ }\mathrm{s.e.}(\log \widehat{RR}) \big).
	\ee

The delta method relies on the approximate linearity represented by the Taylor approximation and approximate normality of the distribution of the maximum likelihood estimates. In particular, the delta method will not perform well when the sampling distribution for $\log \widehat{RR}$ is skewed, which will be a particular concern for large values of $\widehat{RR}$, as the sampling distribution of $\hat{p}_C$ is bounded below by zero.

\paragraph{Bootstrap Method} 
Our bootstrap procedure attempts to reflect the structure of the climate model outputs in the resampling procedure that produces bootstrapped datasets. To generate a bootstrap dataset, we first resample with replacement from the set of ensemble members, as the ensemble members are independent realizations of the climate state. In addition, for each resampled ensemble member, we resample years with replacement from the years represented in the dataset. This second type of resampling is a block bootstrap that is justified by the low correlation in seasonal climate from year to year. Note that by resampling both ensemble members and years, we reduce the discreteness in approximating the sampling distribution that would occur from only resampling from the small number of ensemble members. However, note that in our example, results were similar when either excluding or including the resampling of years.

By repeating the resampling procedure, we produce bootstrap datasets, $D_1, \cdots, D_B$ where $B$ is the bootstrap sample size, e.g., $B=500$. For example, for the actual scenario, we resample with replacement from the five ensemble members and with replacement from the 112 years and the associated smoothed global temperature values. We obtain bootstrap samples with analogous resampling for the counterfactual scenario. The return levels, $\hat z_A^{(1)}, \hat z_A^{(2)}, \cdots, \hat z_A^{(B)}$, are computed from the bootstrapped samples for the actual scenario for the fixed probability $\hat{p}_O$. Pairing each bootstrapped return level estimate from the actual scenario with a bootstrapped dataset from the counterfactual scenario, we obtain bootstrapped probabilities $\hat p_C^{(1)}(\hat z_A^{1}), \hat p_C^{(2)}(\hat z_A^{2}), \cdots, \hat p_C^{(B)}(\hat z_A^{B})$ of exceeding the bootstrapped return levels. We can then calculate $\log\widehat{RR}^{(1)}, \log\widehat{RR}^{(2)}, \cdots, \log\widehat{RR}^{(B)}$, which allows us to estimate the sampling distribution of $\log \widehat{RR}$. From this, one can obtain a bootstrap standard error or confidence interval for the $\log RR$ via standard methods. For the basic bootstrap confidence interval of $\log RR$, we use the 2.5th and 97.5th percentiles of the bootstrapped values for $\log \widehat{RR}^{(b)}, b=1,\ldots,B$, to compute the 95\% confidence interval:
	\be
	\big(  \log \widehat{RR}-(\log \widehat{RR}^{(b)}_{.975}- \log \widehat{RR}),  \mbox{ } \log \widehat{RR}-(\log \widehat{RR}^{(b)}_{.025}- \log \widehat{RR}) \big).
	\ee



\paragraph{Method of Inverting a Likelihood Ratio Test}



The delta method fails when $\hat p_C=0$ ($\widehat{RR}=\infty$) as it relies on asymptotic normality, and the bootstrap method fails for $\hat p_C=0$ and can fail to varying degrees when $\hat p_C$ is very small and one obtains $\log(\widehat{RR})^{(b)}=\infty$ for one or more bootstrap samples. \cite{Hansen_et_al:2014} discussed the case of $\hat p_C=0$ under the counterfactual scenario in the context of event attribution and suggested a one-sided confidence interval  for attributable risk using stationary Poisson processes in the setting where probabilities are estimated simply by empirical proportions. 
Here we propose a likelihood ratio test-based method to find a lower bound for $RR$ that can be employed when extreme value statistics are used. We note that a lower bound is actually more relevant for making an attribution statement than a point estimate of $RR$ as it encapsulates both the potential magnitude of the risk ratio and our uncertainty in estimating it.

A standard approach to finding a confidence interval is to invert a test statistic \citep{Case:Berg:2002}. The basic intuition is that for a hypothesized parameter value, $\theta_{0}$, if we cannot reject the null hypothesis that $\theta=\theta_{0}$ based on the data, then that $\theta_{0}$ is a plausible estimate for the true value of $\theta$ and should be included in a confidence interval for $\theta$. A confidence interval is then constructed by taking all values of $\theta_{0}$ such that a null hypothesis test of $\theta=\theta_{0}$ is not rejected. 

The likelihood ratio test (Sections 9.2.1 \& 10.3.1, \cite{Case:Berg:2002}) compares the likelihood of the data based on the MLE (i.e., the maximized likelihood estimate) to the likelihood of the data when restricting the parameter space (which in the notation above can be expressed as setting $\theta=\theta_{0}$). If the null hypothesis is true then as the sample size goes to infinity, twice the log of the ratio of these two likelihoods has a chi-square distribution with $\nu$ degrees of freedom. $\nu$ is equal to the difference in the number of parameters when comparing the original parameter space to the restricted space. The hypothesis test of
$\theta=\theta_{0}$ is rejected when twice the log of the likelihood ratio exceeds the $1-\alpha$ quantile of the chi-square distribution, which would be the 95th percentile (i.e., $\alpha=0.05$) for a 95\% confidence interval. 

Specifically, we are interested in the plausibility of $RR = \frac{p_{A}}{p_C}=r_0$
versus the alternative that $RR=\frac{p_{A}}{p_C}>r_0$ where $r_0$ is a non-negative constant, so it would be natural
to derive a one-sided confidence interval, $RR\in(RR_{L},\infty)$,
that gives a lower bound, $RR_{L}$, on the risk ratio. The likelihood
ratio test we use here is one where the restricted parameter space
sets $RR=r_{0}$. Under this null hypothesis, which is equivalent to $p_{C} = p_A / r_0$,
we construct the constrained likelihood function by letting $\beta_{0_A} $, $\beta_{1_A}$, $\sigma_A$, $\xi_A$, $\sigma_C$ and $\xi_C$ be free parameters and setting 
\[  \mu_{C}=z_A(\beta_{0_A},\beta_{1_A},\sigma_A,\xi_A)+\frac{\sigma_C}{\xi_C}\big\{ 1-\big(-\log(1-p_A/r_0) \big)^{-\xi_C} \big\},
\]
where $z_A$ 
is the return level corresponding to probability of exceedance under the actual scenario and $p_A$ is based on $\hat p_O$ or chosen in advance without directly making use of the observations. This likelihood ratio test has one degree of freedom, corresponding to the restriction on $\mu_{N}$ in the constrained likelihood.
The joint likelihood for the model output from both the actual scenario and counterfactual scenario can be expressed as 
\begin{eqnarray*}
 L(\theta_A, \theta_C)  & \propto & \exp \bigg\{ -\frac{1}{n_y{_A}} \sum_{i=1}^{n_A} \bigg[ 1+\xi \bigg(\frac{u-\mu_{t_i A}}{\sigma_A}\bigg)\bigg]^{-1/\xi_A}_+ \bigg\} \prod_{i=1}^{m_A} \sigma_A^{-1} \bigg[ 1+\xi_A \bigg(\frac{x_i-\mu_{t_i A}}{\sigma_A}\bigg)\bigg]^{-1/\xi_A-1}_+ \\
&\times & \exp \bigg\{ -\frac{n_C}{n_y{_C}} \bigg[ 1+\xi_C \bigg(\frac{u-\mu_{C}}{\sigma_C}\bigg)\bigg]^{-1/\xi_C}_+ \bigg\} \prod_{j=1}^{m_C} \sigma_C^{-1} \bigg[ 1+\xi_C \bigg(\frac{x_j-\mu_{C}}{\sigma_C}\bigg)\bigg]^{-1/\xi_C-1}_+
\end{eqnarray*}
where $m_A$ is the number of exceedances (out of the total of $n_A$ observations) for the actual scenario and $m_C$ the analogous quantity for the counterfactual scenario. Thus, the lower bound of $RR^L=\min RR$ is found by finding the smallest value $r_0$ such that 
\be \label{eqn:lrt}
2[\log L(\hat\beta_{0_A},\hat\beta_{1_A},\hat\sigma_A,\hat\xi_A,\hat\mu_{C},\hat\sigma_C,\hat\xi_C;x)-\log L(\hat\beta_{0_A},\hat\beta_{1_A},\hat\sigma_A,\hat\xi_A,\hat\sigma_C,\hat\xi_C;x,RR=r_0)]<3.841,
\ee
where $3.841$ is the 95th percentile of a chi-square distribution with one degree of freedom.

Numerically this can be solved by one dimensional minimization subject
to the constraint for the condition (\ref{eqn:lrt}). 
The simplest way to do this is
to move the constraint into the objective function and minimize an
unconstrained problem. The new unconstrained objective function is
\[
r_{0}+c \cdot I(\lambda(r_{0})>3.841)
\]
where $c$ is set to be a large number (mathematically $c=\infty$), $\lambda(\cdot)$ is twice the log of the likelihood ratio, 
and $I(\cdot)$ is an indicator function that evaluates to one if
the inequality is satisfied and zero if not. The resulting objective
function is not continuous, hence many standard optimization techniques
are not applicable. One that can be used here is ``golden section search"
(particularly if the objective function is modified slightly to be
unimodal -- albeit still discontinuous). In \texttt{R}, we use
the \texttt{optimize} function. This function is designed 
for continuous objective functions as it combines golden section search
with parabolic interpolation, but it seems to work reasonably well in
our analyses. 



%% file: result.tex


In this section we apply our proposed methodology to the central US heatwave event. 
The analysis relies on estimation of the probabilities $p_O$ and $p_C$ and the adjusted event magnitude $z_A$. As described in the previous section, we use the smoothed global mean temperature anomaly as a covariate to account for non-stationarity in temperature extremes in both the observations and the model output under the all forcings scenarios. The smoothed global mean temperature anomalies are plotted on Figure \ref{f:ts_all}. Table \ref{tb:bs} gives the parameter estimates from fitting the PP model to observations and to the model output from both scenarios. Note that the estimated shape parameters ($\hat \xi$) are all negative, indicating that the fitted distributions are bounded. 

As shown in Table \ref{tb:bs}, the estimated probability, $\hat{p}_O$, of exceeding the observed extreme value of 2.467 is 0.032. Following the proposed quantile bias correction method, we set $\hat{p}_A=0.032$ and, based on the fitted PP model for the actual scenario, estimate the return level as $\hat z_A=4.842$. Then, using the fitted PP model for the counterfactual scenario, the estimated probability, $\hat p_C$, of an event as or more extreme than $z_C=4.842$  is 1.5e-08. 
The corresponding estimated logarithm of risk ratio is 21.0 (or $RR \approx 2,100,000$), indicating a very large increase in probability of a heatwave due to human influence. Figure \ref{f:CDFs} graphically illustrates the quantile bias correction methodology for this particular case study. Without the bias correction, one would obtain $\hat {p}_A=0.657$ and $\hat {p}_C=0.132$, giving an estimated $RR$ of approximately $5$, which is quite different than the estimate with the bias correction. Note that the observed event is not extreme in the model simulations under the actual scenario, which suggests that without bias correction we would be inappropriately be estimating a $RR$ from a different part of the distribution than is of interest based on the observations.

The uncertainty in estimating $RR$ with the quantile bias correction is quantified using three methods: the delta method, the bootstrap, and our suggested likelihood ratio test-based interval; Table \ref{tb:RR} shows 95\% confidence intervals for $\log RR$ from each method. 
 As discussed in Section 3c, both the delta method and the bootstrap face difficulties when the estimated probability under counterfactual scenario is near zero, as it is here. In this example, the bootstrap resamples often produce estimates of large return levels under the actual scenario that correspond to estimating probabilities of zero under counterfactual scenario. The result is that many of the bootstrap datasets (246 of the 500)  have estimates of $\log RR$ that are infinity, but these bootstrap estimates cannot be sensibly included in the estimate of the bootstrap confidence interval. Hence, the confidence interval in Table \ref{tb:RR}  is calculated based only on the finite values, but we cannot expect this to provide a reliable estimate of the uncertainty.
 
Instead, we focus on the likelihood ratio-based interval described in the previous section. We apply our method by inverting a LRT in two ways. First we ignore uncertainty in $\hat{z}_A$ and consider only uncertainty in $\hat p_C$, and second we consider uncertainty in both $\hat z_A$ and $\hat p_C$ (note that when we consider only uncertainty in $\hat p_C$, one can derive a LRT-based interval analogously to that derived in Section 3c). 

The estimated lower bound, when considering both sources of uncertainties, is 4.0 (i.e., 16.1 on the original scale of the risk ratio), which indicates strong evidence that the true risk ratio is substantially elevated under actual scenario compared to counterfactual scenario. As expected, the lower bound is lower (4.0) when considering both sources of uncertainty than when considering only uncertainty in $\hat p_C$ (4.3). 

In Section 3c, we argued that a precise event magnitude and corresponding $p_O$ is not necessary to making confident event attribution statements. Rather, the sensitivity of the risk ratio to a plausible range of extreme event definitions is essential. 
Table \ref{tb:sensitivity} shows the sensitivity of the risk ratio and its lower bound to various values of ${p}_O={p}_A$. Critically, while the estimate of the risk ratio varies dramatically as one varies the event definition, with the estimated risk ratio as large as infinity, the lower bound from the one-sided confidence interval is quite stable for a wide range of event definitions. This is a critically important component to the confident event attribution statement: ``For the summer 2011 central US heat wave, anthropogenic changes to the atmospheric composition caused the chance of the observed temperature anomaly to be increased by \textit{at least} a factor of 16.1."  Of course this statement is conditional on the climate model accurately representing relative changes in probabilities of extreme events under the different scenarios after the quantile-based correction.

%% file: conclusion.tex

We present an approach to extreme event attribution that addresses differences in the scales of variability between observations and model output using the methodology of quantile-based bias correction in the context of a formal statistical treatment of uncertainty. The correction rescales matching quantiles between the observations and the models to obtain an event in realistically-forced climate model simulations of corresponding rarity to the actual extreme weather or climate event of interest. We develop a procedure for estimation and for quantifying uncertainty in the risk ratio, a measure of the anthropogenic effect on the change in the chances of an extreme event. In particular we calculate a lower bound on the risk ratio by inverting a likelihood ratio test statistic that can be used even when the estimated probability of the event is zero or near-zero in climate model simulations of a hypothetical world without anthropogenic climate change. This lower bound provides the key element in constructing confident attribution statements about the human influence on individual extreme weather and climate events. 

We caution that bias correction can mask serious errors and is not a replacement for expert judgment and physical insight into the source of the bias between model and observation. For instance in our case study, it is well known that extreme temperatures in Texas and Oklahoma are associated with the La Ni\~{n}a phase of ocean surface temperatures. The statistical methods presented here could account for this source of bias by including an El Ni\~{n}o/La Ni\~{n}a index as a covariate in the statistical model for event probabilities in the model dataset (see Section 3b) and bias correct the index rather than directly bias correcting the distribution for the variable of interest. Pursuing such ideas is beyond the scope of our work here but could lead to an approach that offers more insight into the source of bias and provide a physically-based justification for the bias correction.

The lower bound on the risk ratio estimated using our proposed method implies a substantial increase in the probability of reaching or exceeding the observed extreme temperature of 2011 central US heatwave event under human-influenced climate change. However the precise probability and magnitude of the observed extreme event is not a key component in extreme event attribution analyses. We explored the sensitivity of the lower bound of the risk ratio to various definitions of the event (i.e., probabilities corresponding to different magnitudes of extreme events) and found that the lower bound of the risk ratio confidence interval is more stable than point estimates of the risk ratio. As a result, confident attribution statements about the minimum amount of anthropogenic influence on extreme events are more readily constructed than statements about the most likely amount of anthropogenic influence. We also maintain that such more conservative statements are more consistent with the vast literature of attribution statements about the human influence on trends in the average state of the climate.  


%% file: figures_and_tables.tex

\begin{figure}
  \begin{center}
  \includegraphics{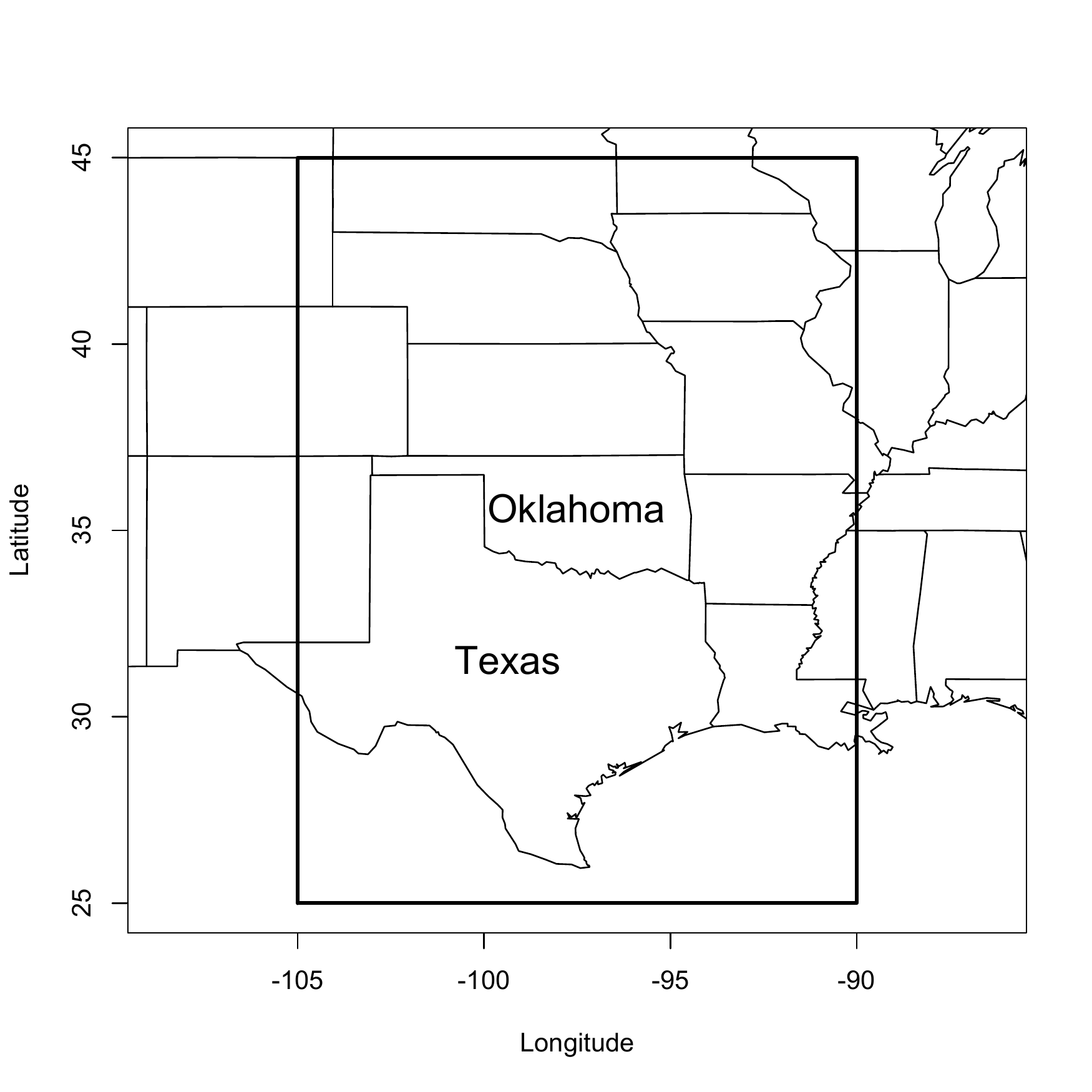}
  \caption{Central United States region, 90$^{\circ}$W to 105$^{\circ}$W in longitude and 25$^{\circ}$N to 45$^{\circ}$N in latitude (bold rectangular area), covering the states of Texas and Oklahoma.}
 \label{f:map}
 \end{center}
\end{figure}

\begin{figure}
  \begin{center}
  \includegraphics{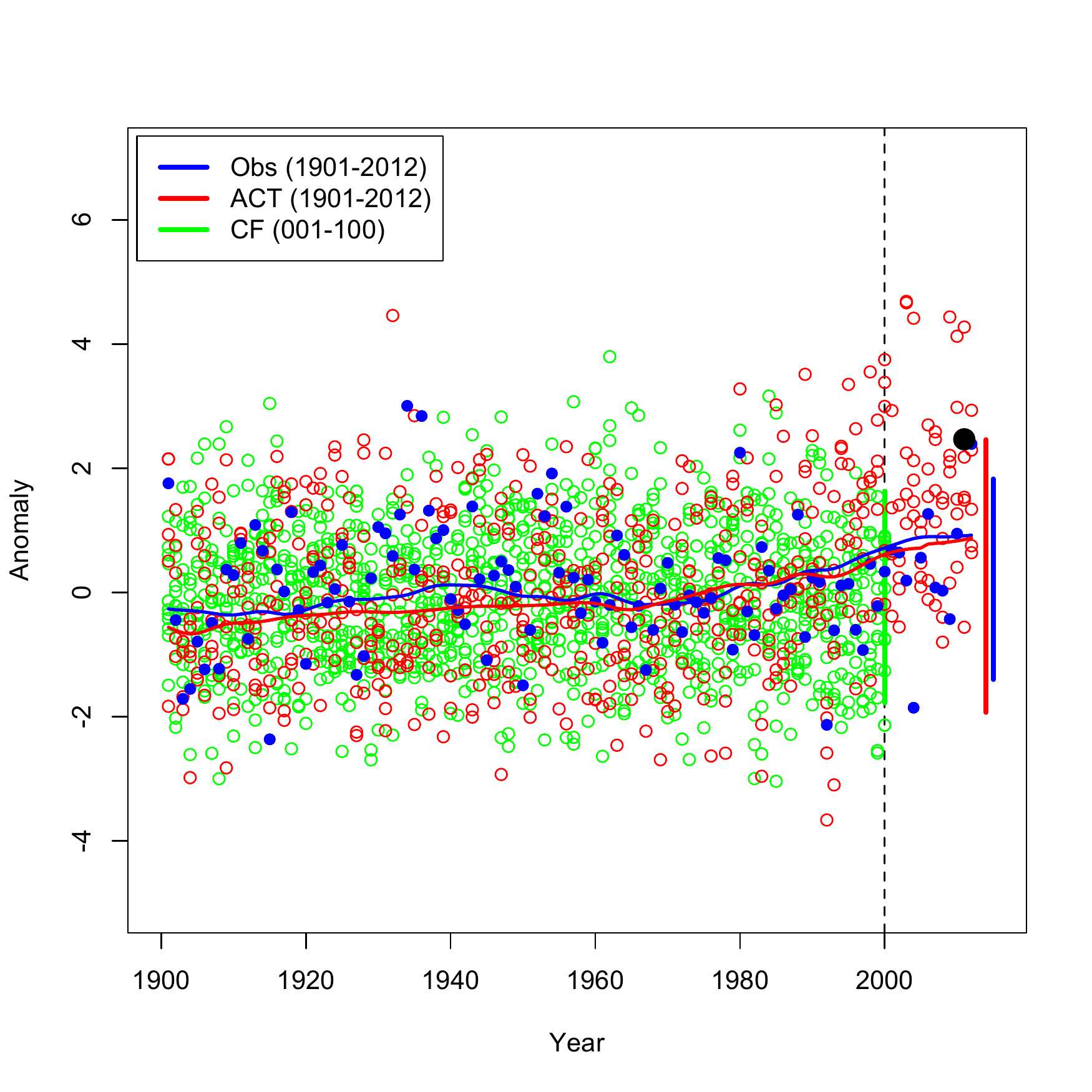}
  \caption{Illustration of the mismatch in scales between observations
    and model output for central US summer temperatures. Observed values
    for 1901-2012 (blue), model output under actual scenario for
    1901-2012 (red) and model output under counterfactual scenario for
    100-year time period (green). The vertical lines show the 5-95\% range of
    values for the different datasets. The larger black dot represents
    the observed value of 2.467 for 2011. The blue and red lines
    represent smoothed global mean temperature anomalies used as
    observational and actual scenario model output covariates, respectively.}
 \label{f:ts_all}
 \end{center}
\end{figure}

\begin{figure}
  \begin{center}
  \includegraphics{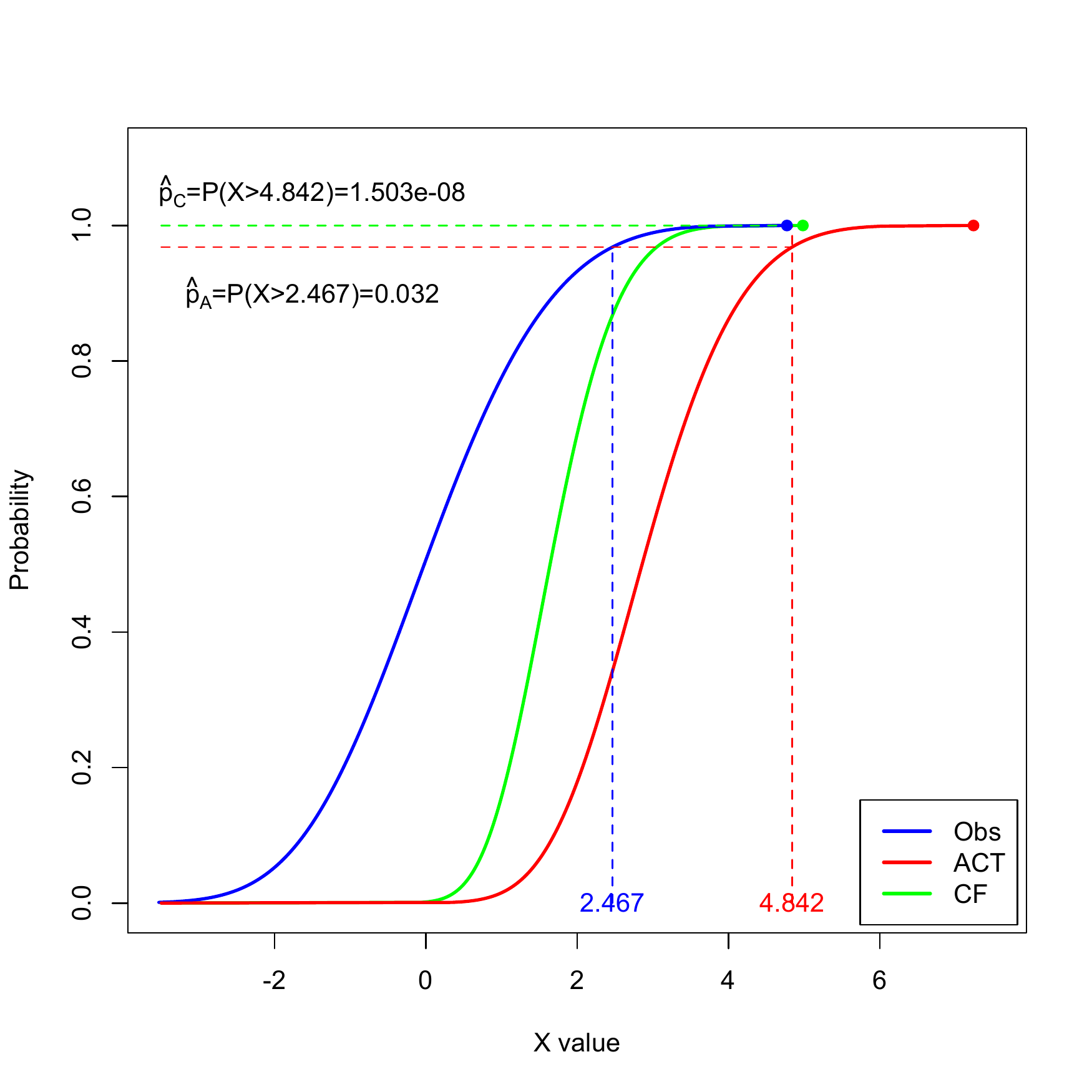}
  \caption{Demonstration of the quantile bias correction applied to the central US heatwave example, showing estimated cumulative distribution functions of observed (blue) and modeled datasets under actual scenario (red) and counterfactual scenario (green). The blue dashed line shows the observed event, with the horizontal red dashed line translating the observed event magnitude to the equivalent magnitude under the actual scenario, holding $\hat{p}_O=\hat{p}_A$. For the event magnitude indicated by the vertical red dashed line, the green dashed line indicates the probability under counterfactual scenario. The three colored dots represent the upper bounds of each distribution function, which occurs because with a negative shape parameter (as is estimated in these cases), the extreme value distribution has a finite upper bound.
}
 \label{f:CDFs}
 \end{center}
\end{figure}

\begin{figure}
  \begin{center}
   \subfigure[]{\includegraphics[width=0.32\textwidth]{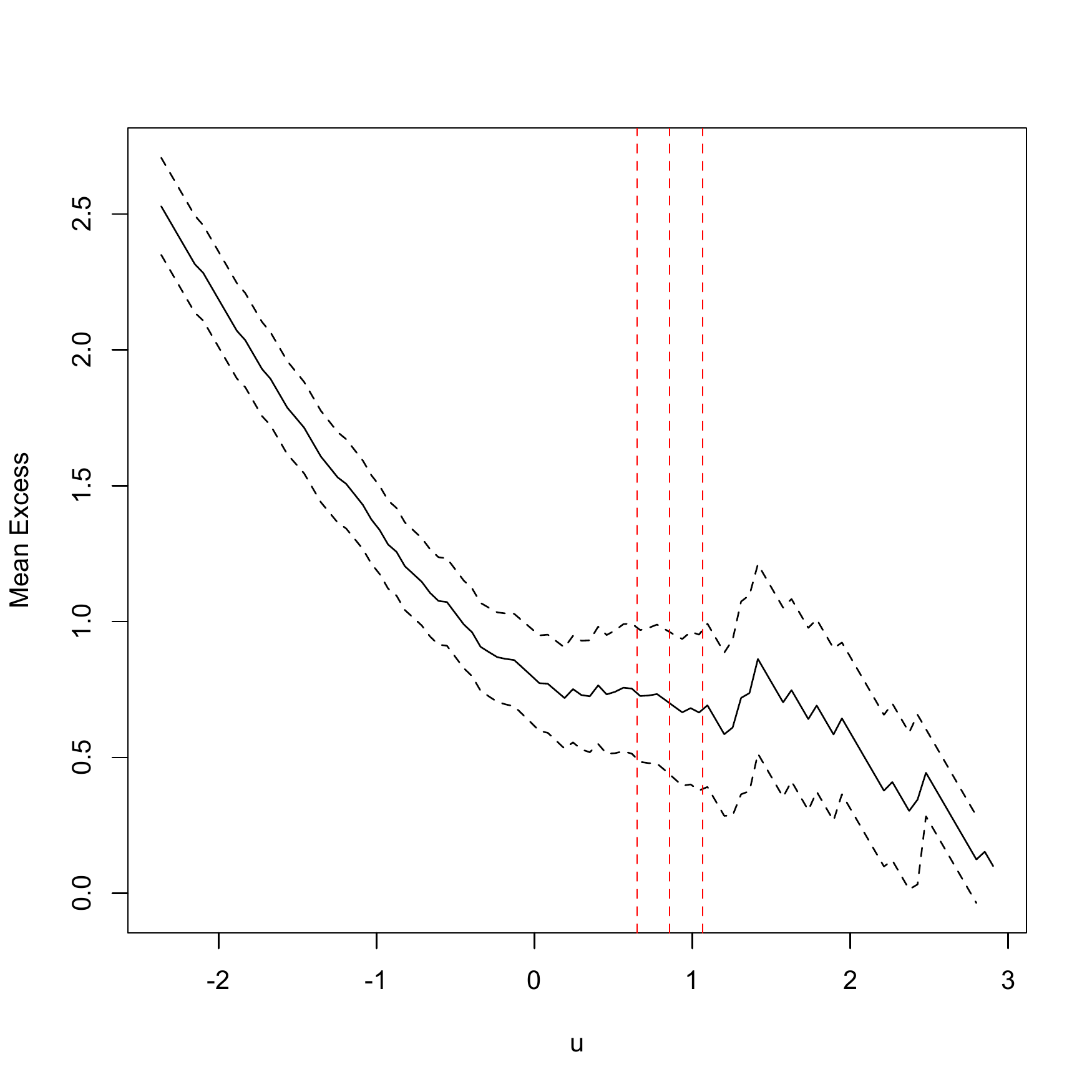}}
   \subfigure[]{\includegraphics[width=0.32\textwidth]{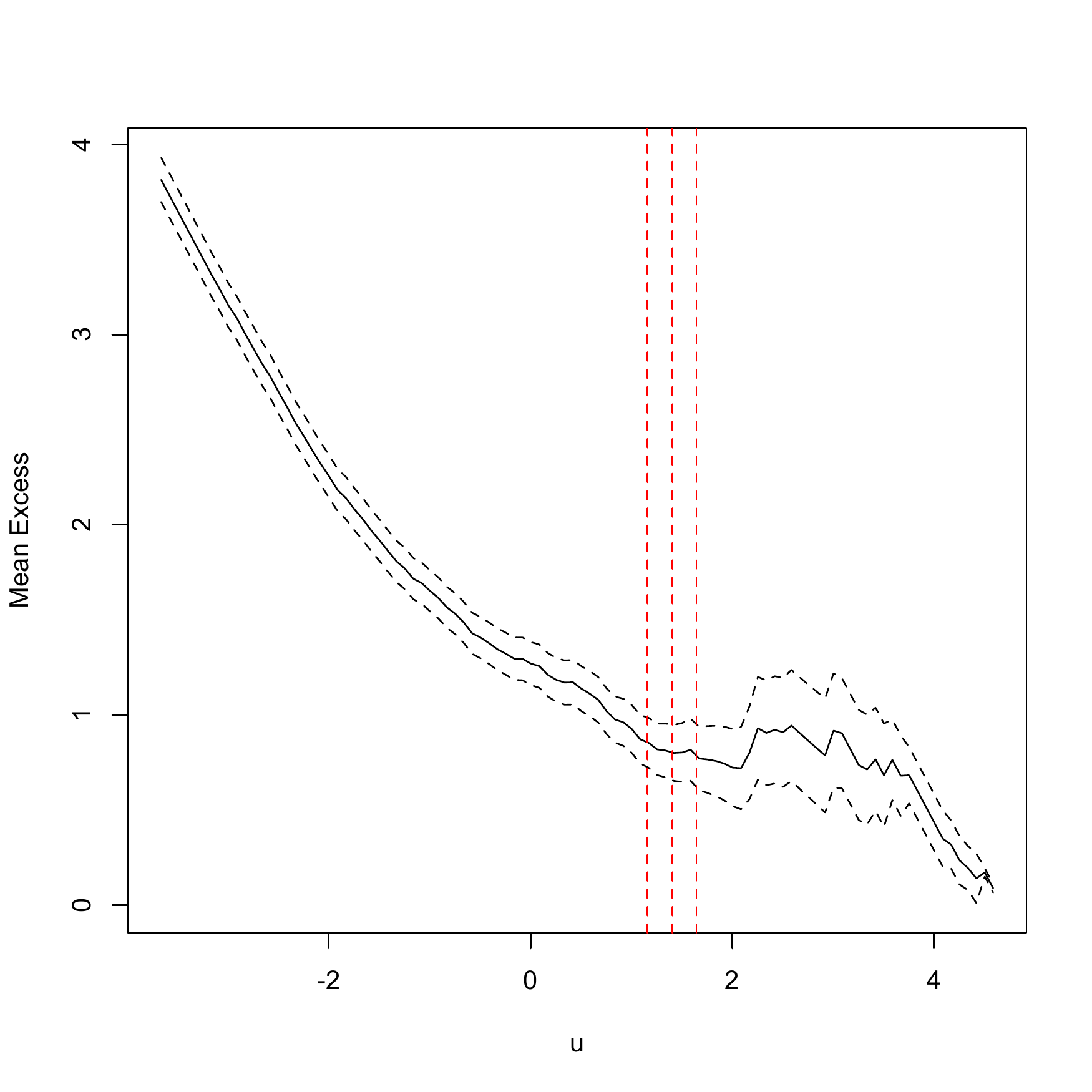}}
   \subfigure[]{\includegraphics[width=0.32\textwidth]{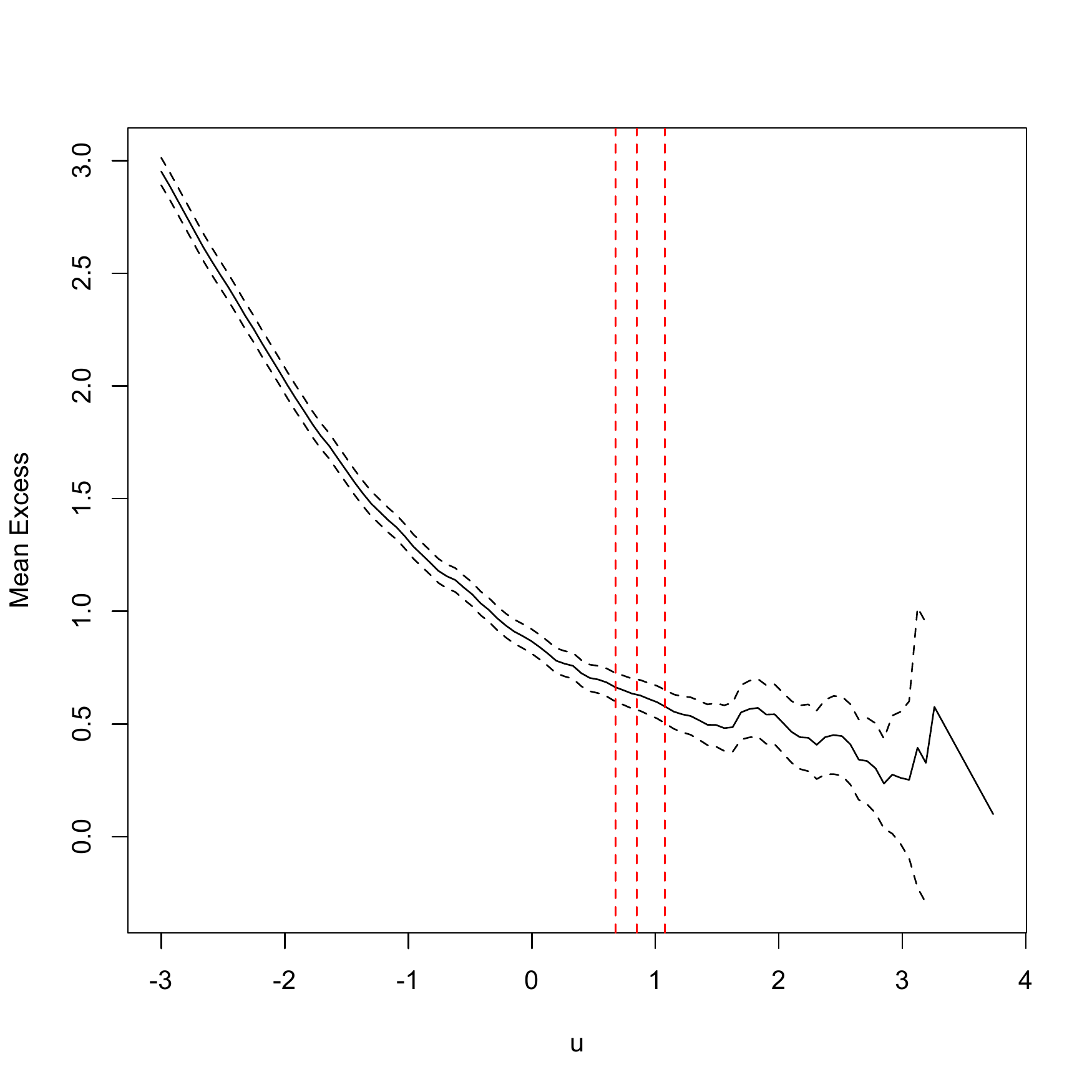}}
  \caption{Mean residual life plot for each dataset: (a) observations, (b) model output under actual scenario, and (c) model output under counterfactual scenario. Red dashed lines represent the 75th, 80th, and 85th percentiles, respectively, as possible choices of thresholds. We chose the 80th percentile as a reasonable threshold beyond which there are relatively linear trends.}
 \label{f:mrlp}
 \end{center}
\end{figure}


\begin{table}[t]
\caption{Parameter estimates from the point process model fitted to
  observations (top, 1901-2012), actual scenario model output
  (middle, 1901-2012), and counterfactual scenario model output
  (bottom, 100 years). The right column gives the estimated return
  levels and/or probabilities calculated in the steps of the quantile
  bias correction method. The threshold, $u$, is  the 80th percentile of values for each given dataset.}\label{tb:bs} 
\begin{center}

\begin{tabular}{l|cccc|c}
\hline\hline
Observation & location& & scale & shape &\\
$u=0.856$ & $\hat{\beta}_0$ & $\hat{\beta}_1$ & $\hat{\sigma}$ & $\hat{\xi}$ & $\hat{p}_O=P(Z>2.467$)\\
\hline
global mean tmp & -0.802 & 0.404 & 1.250 & -0.239 & 0.032\\
\hline
\multicolumn{1}{c}{}  & \multicolumn{1}{c}{}  & \multicolumn{1}{c}{} & \multicolumn{1}{c}{} & \multicolumn{1}{c}{} & \\
\hline\hline
Model  (actual scenario)& location& & scale & shape & \\
$u=1.405$ & $\hat{\beta}_{0_A}$ & $\hat{\beta}_{1_A}$ & $\hat{\sigma}_A$ & $\hat{\xi}_A$  & $\hat{z}_A$ \\
\hline
global mean tmp & 1.263 & 1.382 & 0.926 & -0.197  & 4.842\\
\hline
\multicolumn{1}{c}{}  & \multicolumn{1}{c}{}  & \multicolumn{1}{c}{} & \multicolumn{1}{c}{} & \multicolumn{1}{c}{} & \\
\hline\hline
Model (counterfactual) & location& & scale & shape &\\
$u=0.811$ & $\hat{\mu}_{C}$ &  & $\hat{\sigma}_C$ & $\hat{\xi}_C$ & $\hat{p}_C=P(Z>4.842)$ \\
\hline
no trend ($K=0$) & 1.415 &  & 0.638 & -0.179 &  1.503e-08 \\
\hline
\end{tabular}

\end{center}
\end{table}

\begin{table}[t]
\caption{Estimated $\log RR$ and corresponding confidence intervals
  using delta method, bootstrap resampling (B=500), and the proposed
  likelihood ratio test (LRT)-based method giving a lower bound for the
  risk ratio. For the bootstrap, 246 of the 500 bootstrap samples are
  excluded as the bootstrapped $RR$ estimate is infinity. For the LRT-based approach, we consider two cases of uncertainty quantification: first uncertainty only in estimating $p_C$, and second uncertainty in estimating both $z_A$ and $p_C$.}\label{tb:RR} 
\begin{center}
\begin{tabular}{l|c}
\hline\hline
$\log_2 \widehat{RR}$ & {\bf 21.0}\\ 
\hline
\multicolumn{1}{c}{}  & \multicolumn{1}{c}{}  \\
\multicolumn{1}{l}{\bf Delta method} &  \multicolumn{1}{c}{}\\
\hline
& [16.8, 25.2]\\
\hline

\multicolumn{1}{c}{}  & \multicolumn{1}{c}{} \\
\multicolumn{1}{l}{\bf Bootstrap method} &  \multicolumn{1}{c}{}  \\
\hline
& [12.2, 39.4]  \\
\hline

\multicolumn{1}{c}{}  & \multicolumn{1}{c}{} \\
\multicolumn{1}{l}{\bf LRT-based method} &  \multicolumn{1}{c}{}  \\
\hline
UQ for $\hat p_C$ & [4.3, $\infty$)  \\
\hline
UQ for $\hat z_A$ and $\hat p_C$ & [4.0, $\infty$)  \\
\hline
\end{tabular}

\end{center}
\end{table}

\begin{table}[t]
\caption{Sensitivity of results to definition of the event, i.e., different values of $p_O=p_A$.}\label{tb:sensitivity} 
\begin{center}
\begin{tabular}{c|cc|c|c|c}
\hline\hline
$p_A$ & $\hat z_A$ & $\hat p_C$ & $\log_2 \widehat{RR}$ & one-sided CI for $\log_2 RR$ & lower bound of $RR$\\
$p_O=p_A$ & & & & $(\alpha=.05)$ &  \\
\hline
0.200 & 3.7 & 2.8e-03 & 6.1 & [3.0, $\infty$) & 8.0\\
0.100 & 4.2 & 1.9e-04 & 9.1 & [3.6, $\infty$) & 11.7\\
0.050 & 4.6 & 3.1e-06 & 14.0 & [3.9, $\infty$) & 14.8\\
0.032 & 4.8 & 1.5e-08 & 21.0 & [4.0, $\infty$) & 16.1\\
0.023 & 5.0 & 0 & $\infty$ & [4.1, $\infty$) & 16.8\\
0.010 & 5.3 & 0 & $\infty$  & [4.1, $\infty$) & 16.9\\


\hline
\end{tabular}

\end{center}
\end{table}

%% file: appendix.tex

Extreme value theory (EVT) provides a statistical theory of extreme values that models the tail of a probability distribution. Univariate extreme value theory to study so-called block maxima (e.g., annual or seasonal maxima of daily data) is well-developed. The theory shows that the distribution of the maxima converges to a distribution function $G$, 
    \be  \label{eqn:gev}
       G(x;\mu,\sigma,\xi)=\exp \bigg\{-\bigg(1+\xi \frac{x-\mu}{\sigma}\bigg )_+ ^{-1/\xi}\bigg\}, \quad (x_+ =\max (0,x))
    \ee
that is known as the generalized extreme value (GEV) distribution. The  parameters $\mu$, $\sigma$, and $\xi$ are known as the location, scale and shape parameters, respectively.   The shape parameter, $\xi$, determines the type of tail behavior --- whether the tail is heavy ($\xi>0$), light ($\xi \rightarrow 0$), or bounded ($\xi<0$), implying a short-tailed distribution. For example, analysts usually obtain a negative estimated shape parameter for temperature data and a non-negative estimated shape parameter for precipitation data.

Return levels are quantiles --- a return level $z$ such that $P(Z>z)=p$ implies that the level $z$ is expected to be exceeded once every $1/p$ years on average. The probability $p$ of exceeding $z$ is easily obtained in closed form, given  $\mu$, $\sigma$, and $\xi$, based on the distribution function (\ref{eqn:gev}),  
	\be \label{eqn:return_prob}
	p=1-P(Z \leq z)=1-\exp \bigg\{-\bigg(1+\xi \frac{z-\mu}{\sigma}\bigg )_+ ^{-1/\xi}\bigg\}.
	\ee
As a counterpart to this, given $p$, the return level is obtained by solving the equation $P(Z>z)=p$, which gives
	\be \label{eqn:return_value}
	z=\mu-\frac{\sigma}{\xi}\big\{ 1-\big(-\log(1-p) \big)^{-\xi} \big\} \quad (\xi \neq 0).
	\ee 

However, the block maxima approach only uses the maximum (or analogously the minimum when analyzing extreme low values) of blocks in time series data. An alternative that can make use of more of the data is the peaks over threshold (POT) approach \citep{Coles:2001, Katz:et:al2002}. POT modeling is based on the observations above a high threshold, $u$. The distribution of exceedances over the threshold is approximated by a generalized Pareto distribution (GPD) as $u$ becomes sufficiently large.
In this approach, the limiting distribution of threshold exceedances is characterized by the following: for $x>u$, 
	\be
	P(X \leq x|X>u) = 1-\bigg( 1+\xi \frac{x-u}{\sigma_u} \bigg)_+^{-1/\xi}.
	\ee 
The scale parameter $\sigma_u>0$ depends on the threshold. As with the GEV distribution the shape parameter, $\xi$, determines the tail behavior. 
   
The point process (PP) provides a closely-related alternative peaks over threshold approach to the GPD that is convenient because the PP parameters can be directly related to the GEV parameters and then the GEV equations above can be used to calculate return values and return probabilities. The corresponding likelihood of the threshold excesses can be approximated by a Poisson distribution with the intensity measure depending on $\mu$, $\sigma$, and $\xi$, where $\mu$, $\sigma$, and $\xi$ are location, scale, and shape parameters equivalent to those in the GEV distribution (\ref{eqn:gev}). More precisely, for a vector of $n$ observations $X_1, X_2, \cdots, X_n$ standardized under the conditions of GEV distribution, the point process on regions of $(0,1) \times [u,\infty)$ converges to a Poisson process with the intensity measure given by
	\be
	\Lambda \big([t_1, t_2] \times (x,\infty) \big)=(t_2-t_1) \bigg [ 1+\xi \bigg(\frac{x-\mu}{\sigma}\bigg) \bigg ]_+^{-1/\xi}.
	\ee
Taking $m$ to be the number of observations above the threshold $u$ (out of the total of $n$ observations), the likelihood function is 
	\be \label{eqn:pp_lh} 
	L(\theta;x_1, x_2, \cdots, x_n) \propto \exp \bigg\{ -\frac{n}{n_y} \bigg[ 1+\xi \bigg(\frac{u-\mu}{\sigma}\bigg)\bigg]^{-1/\xi}_+ \bigg\} \prod_{i=1}^{m} \sigma^{-1} \bigg[ 1+\xi \bigg(\frac{x_i-\mu}{\sigma}\bigg)\bigg]^{-1/\xi-1}_+
	\ee
where $n_y$ is number of observations per year (e.g., $n_y=5$ for the all forcings ensemble and $n_y=12$ for the counterfactual ensemble).

%% file: paper_SJ-CP-MW.bbl
\begin{thebibliography}{32}
\providecommand{\natexlab}[1]{#1}
\providecommand{\url}[1]{\texttt{#1}}
\providecommand{\urlprefix}{URL }
\expandafter\ifx\csname urlstyle\endcsname\relax
  \providecommand{\doi}[1]{doi:\discretionary{}{}{}#1}\else
  \providecommand{\doi}{doi:\discretionary{}{}{}\begingroup
  \urlstyle{rm}\Url}\fi
\providecommand{\eprint}[2][]{\url{#2}}

\bibitem[{Allen(2003)}]{Allen:2003}
Allen, M.~R., 2003: Liability for climate change. \textit{Nature},
  \textbf{421~(6926)}, 891--892.

\bibitem[{Bindoff et~al.(2013)}]{Bindoff:2013}
Bindoff, N., et~al., 2013: \textit{Detection and Attribution of Climate Change:
  from Global to Regional In: Climate Change 2013: The Physical Science Basis.
  Contribution of Working Group I to the Fifth Assessment Report of the
  Intergovernmental Panel on Climate Change [Stocker, T.F., D. Qin, G.-K.
  Plattner, M. Tignor, S.K. Allen, J. Boschung, A. Nauels, Y. Xia, V. Bex and
  P.M. Midgley (eds.)]}. Cambridge University Press, Cambridge, United Kingdom
  and New York, NY, USA.

\bibitem[{Casella and Berger(2002)}]{Case:Berg:2002}
Casella, G. and R.~L. Berger, 2002: \textit{Statistical {I}nference}. 2d ed.,
  Thomson Learning, Australia Pacific Grove, CA.

\bibitem[{Christidis et~al.(2013)Christidis, Stott, Scaife, Arribas, Jones,
  Copsey, Knight, and Tennant}]{Chrisidis_et_al:2013}
Christidis, N., P.~A. Stott, A.~A. Scaife, A.~Arribas, G.~S. Jones, D.~Copsey,
  J.~R. Knight, and W.~J. Tennant, 2013: A new {HadGEM3-A-Based} system for
  attribution of weather- and climate-related extreme events. \textit{Journal
  of Climate}, \textbf{26}, 2756--2783.

\bibitem[{Coles(2001)}]{Coles:2001}
Coles, S.~G., 2001: \textit{An Introduction to Statistical Modeling of Extreme
  Values}. Springer Verlag, New York.

\bibitem[{Edwards et~al.(2014)Edwards, Bunkers, Abatzoglou, Todey, and
  Parker}]{Edwards_et_al:2014}
Edwards, L.~M., M.~J. Bunkers, J.~T. Abatzoglou, D.~P. Todey, and L.~E. Parker,
  2014: {October 2013 Blizzard in western South Dakota} [in ``{Explaining
  Extremes of 2013 from a Climate Perspective}"]. \textit{Bulletin of the
  American Meteorological Society}, \textbf{95~(9)}, S23--S26.

\bibitem[{Fischer and Knutti(2015)}]{Fischer_Knutti:2015}
Fischer, E.~M. and R.~Knutti, 2015: Anthropogenic contribution to global
  occurrence of heavy-precipitation and high-temperature extremes.
  \textit{Nature Climate Change}, \textbf{5~(6)}, 560--564.

\bibitem[{Furrer et~al.(2010)Furrer, Katz, Walter, and
  Furrer}]{Furrer_et_al:2010}
Furrer, E.~M., R.~W. Katz, M.~D. Walter, and R.~Furrer, 2010: Statistical
  modeling of hot spells and heat waves. \textit{Climate Research},
  \textbf{43}, 191--205.

\bibitem[{Gilleland and Katz(2011)}]{extRemes:2011}
Gilleland, E. and R.~W. Katz, 2011: New software to analyze how extremes change
  over time. \textit{Eos}, \textbf{92~(2)}, 13--14.

\bibitem[{Hansen et~al.(2014)Hansen, Auffhammer, and Solow}]{Hansen_et_al:2014}
Hansen, G., M.~Auffhammer, and A.~R. Solow, 2014: On the attribution of a
  single event to climate change. \textit{J. Climate}, \textbf{27}, 8297--8301.

\bibitem[{Harris et~al.(2014)Harris, Jones, Osborn, and
  Lister}]{Harris_et_al:2014}
Harris, I., P.~D. Jones, T.~J. Osborn, and D.~Lister, 2014: Updated
  high-resolution grids of monthly climatic observations -- the {CRU TS3.10
  Dataset}. \textit{Int. J. Climatol.}, \textbf{34}, 623--642,
  \doi{10.1002/joc.3711}.

\bibitem[{Herring et~al.(2014)Herring, Hoerling, Peterson, and
  Stott}]{BAMS:2013}
Herring, S.~C., M.~P. Hoerling, T.~C. Peterson, and P.~A. Stott, (Eds.) , 2014:
  \textit{Explaining Extreme Events of 2013 from a Climate Perspective}.
  Bulletin of the American Meteorological Society, 95 (9), S1--S96.

\bibitem[{Hoerling et~al.(2013)}]{Hoerling_et_al:2008}
Hoerling, M., et~al., 2013: Anatomy of an extreme event. \textit{J. Climate},
  \textbf{26}, 2811--2832.

\bibitem[{Jaeger et~al.(2008)Jaeger, Krause, Haas, Klein, and
  Hasselmann}]{Jaeger_et_al:2008}
Jaeger, C.~C., J.~Krause, A.~Haas, R.~Klein, and K.~Hasselmann, 2008: A method
  for computing the fraction of attributable risk related to climate damages.
  \textit{Risk Analysis}, \textbf{28~(4)}, 815--823,
  \doi{10.1111/j.1539-6924.2008.01070.x},
  \urlprefix\url{http://dx.doi.org/10.1111/j.1539-6924.2008.01070.x}.

\bibitem[{Katz et~al.(2002)Katz, Parlange, and Naveau}]{Katz:et:al2002}
Katz, R.~W., M.~B. Parlange, and P.~Naveau, 2002: Statistics of extremes in
  hydrology. \textit{Advances in Water Resources}, \textbf{25}, 1287---1304.

\bibitem[{Kharin and Zwiers(2005)}]{Kharin_Zwiers:2005}
Kharin, V.~V. and F.~W. Zwiers, 2005: Estimating extremes in transient climate
  change simulations. \textit{J. Climate}, \textbf{18}, 1156--1173.

\bibitem[{Li et~al.(2010)Li, Sheffield, and Wood}]{Li_et_al:2010}
Li, H., J.~Sheffield, and E.~F. Wood, 2010: Bias correction of monthly
  precipitation and temperature fields from {Intergovernmental Panel on Climate
  Change AR4} models using equidistant quantile matching. \textit{J. of
  Geophys. Res.}, \textbf{115~(D10101)}, \doi{10.1029/2009JD012882}.

\bibitem[{Maurer and Hidalgo(2008)}]{Maurer_et_al:2008}
Maurer, E.~P. and H.~G. Hidalgo, 2008: Utility of daily vs. monthly large-scale
  climate data: an intercomparison of two statistical downscaling methods.
  \textit{Hydrol. Earth Syst. Sci.}, \textbf{12}, 551--563.

\bibitem[{Meehl et~al.(2012)Meehl, Arblaster, and
  Branstator}]{Meehl_et_al:2012}
Meehl, G.~A., J.~M. Arblaster, and G.~Branstator, 2012: Mechanisms contributing
  to the warming hole and the consequent {U.S. East--West} differential of heat
  extremes. \textit{Journal of Climate}, \textbf{25}, 6394--6408.

\bibitem[{Min et~al.(2013)Min, Zhang, Zwiers, Shiogama, Tung, and
  Wehner}]{Min_et_al:2013}
Min, S.-K., X.~Zhang, F.~Zwiers, H.~Shiogama, Y.-S. Tung, and M.~Wehner, 2013:
  Multimodel detection and attribution of extreme temperature changes.
  \textit{Journal of Climate}, \textbf{26}, 7430--7451.

\bibitem[{Pall et~al.(2011)Pall, Aina, Stone, Stott, Nozawa, Hilberts, Lohmann,
  and Allen}]{Pall_et_al:2011}
Pall, P., T.~Aina, D.~A. Stone, P.~A. Stott, T.~Nozawa, A.~G.~J. Hilberts,
  D.~Lohmann, and M.~R. Allen, 2011: Anthropogenic greenhouse gas contribution
  to flood risk in {England and Wales} in autumn 2000. \textit{Nature},
  \textbf{470~(7334)}, 382--385.

\bibitem[{Palmer(1999)}]{Palmer:1999}
Palmer, T.~N., 1999: A nonlinear dynamical perspective on climate prediction.
  \textit{J. Climate}, \textbf{12}, 575--591.

\bibitem[{Panofsky and Brier(1968)}]{Panofsky_Brier:1968}
Panofsky, H.~A. and G.~W. Brier, 1968: \textit{Some Applications of Statistics
  to Meteorology}. The Pennsylvania State University, University Park, PA, USA,
  224 pp.

\bibitem[{Peterson et~al.(2013)Peterson, Hoerling, Stott, and
  Herring}]{BAMS:2012}
Peterson, T.~C., M.~P. Hoerling, P.~A. Stott, and S.~Herring, (Eds.) , 2013:
  \textit{Explaining Extreme Events of 2012 from a Climate Perspective}.
  Bulletin of the American Meteorological Society, 94 (9), S1--S74.

\bibitem[{Peterson et~al.(2012)Peterson, Stott, and Herring}]{BAMS:2011}
Peterson, T.~C., P.~A. Stott, and S.~Herring, 2012: Explaining extreme events
  of 2011 from a climate perspective. \textit{Bulletin of the American
  Meteorological Society}, \textbf{93}, 1041--1067,
  \doi{10.1175/BAMS-D-12-00021.1}.

\bibitem[{Scarrott and MacDonald(2012)}]{Scarrott_MacDonald:2012}
Scarrott, C. and A.~MacDonald, 2012: A review of extreme value threshold
  estimation and uncertainty quantification. \textit{REVSTAT - Statistical
  Journal}, \textbf{10~(1)}, 33--60.

\bibitem[{Smith(1989)}]{Smith:1989}
Smith, R.~L., 1989: Extreme value analysis of environmental time series: An
  application to trend detection in ground-level ozone (with discussion).
  \textit{Statistical Science}, \textbf{4}, 367--393.

\bibitem[{Solomon et~al.(2007)Solomon, Qin, Manning, Chen, Marquis, Averyt,
  Tignor, and Miller}]{IPCC_AR4}
Solomon, S., D.~Qin, M.~Manning, Z.~Chen, M.~Marquis, K.~B. Averyt, M.~Tignor,
  and H.~L. Miller, 2007: \textit{In Climate Change 2007: The Physical Science
  Basis. Contribution of Working Group I to the Fourth Assessment Report of the
  Intergovernmental Panel on Climate Change}. Cambridge University Press,
  Cambridge, United Kingdom and New York, NY, USA.

\bibitem[{Stone et~al.(2013)Stone, Paciorek, Prabhat, Pall, and
  Wehner}]{Stone_et_al:2013}
Stone, D.~A., C.~J. Paciorek, P.~Prabhat, P.~Pall, and M.~F. Wehner, 2013:
  Inferring the anthropogenic contribution to local temperature extremes.
  \textit{Proc. Natl. Acad. Sci. USA}, \textbf{110~(17)}, E1543.

\bibitem[{Stott et~al.(2004)Stott, Stone, and Allen}]{Stott_et_al:2004}
Stott, P.~A., D.~A. Stone, and M.~R. Allen, 2004: Human contribution to the
  {European} heatwave of 2003. \textit{Nature}, \textbf{432~(7017)}, 610--614.

\bibitem[{Wolski et~al.(2014)Wolski, Stone, Tadross, Wehner, and
  Hewitson}]{Wolski_et_al:2013}
Wolski, P., D.~Stone, M.~Tadross, M.~Wehner, and B.~Hewitson, 2014:
  {Attribution of floods in the Okavango basin, Southern Africa}.
  \textit{Journal of Hydrology}, \textbf{511}, 350--358,
  \doi{10.1016/j.jhydrol.2014.01.055}.

\bibitem[{Wood et~al.(2004)Wood, Leung, Sridhar, and
  Lettenmaier}]{Wood_et_al:2004}
Wood, A., L.~R. Leung, V.~Sridhar, and D.~P. Lettenmaier, 2004: Hydrologic
  implications of dynamical and statistical approaches to downscaling climate
  model outputs. \textit{Climatic Change}, \textbf{62}, 189--216.

\end{thebibliography}
